# Modification Of Gtd From Flat File Format To Olap For Data Mining

Karanjit Singh and Dr. Shuchita Bhasin
HQ Base Workshop Group EME, Indian Army, Computer Science and IT Dept, Kurukshetra University

**Abstract - This document is part of original research work by the authors in a bid to explore new fields for applying Data Mining Techniques. The sample data is part of a large data set from University of Maryland (UMD) and outlines how more meaningful patterns can be discovered by preprocessing the data in the form of OLAP cubes**

## I. INTRODUCTION

Application of Data Mining Tools for Terror Data Mining is a lesser talked about field[1]. Lot of research efforts are going into capturing the data from incident reports in the past and structuring the data for analysis. Unfortunately there are not many sources on the net. One such database available [2] in a single tabular form, is an Open Source Terrorism Incident events Database called Global Terrorism Database (GTD). This covers terrorism incidents around the world from 1970 through 2008 (with continuing annual updates). It includes systematic data on US, as well as transnational and international terrorist incidents that have occurred during this time period and as on now includes more than 87,000 cases. For each GTD incident, information is available on the date and location of the incident, the weapons used and nature of the target, the number of casualties, and--when identifiable--the group or individual responsible. However the format in which it is available lends itself only to limited analysis unless suitable tools for analysis are used. This paper analyses the available data fields and suggests a format for OLAP and subsequent data mining. The data base has been obtained from the National Consortium for the Study of Terrorism and Responses to Terrorism (START) initiative at University of Maryland, from their online interface at http:://www.start.umd.edu/gtd/ in an effort to increase understanding of terrorist violence so that it can be more readily studied and defeated.

## II. CHARACTERISTICS OF AVAILABLE DATA

The main characteristics of the GTD [4] are:-

- Information on over 87,000 terrorist attacks
- Currently the most comprehensive unclassified data base on terrorist events in the world
- Information on more than 38,000 bombings, 13,000 assassinations, and 4,000 kidnappings since 1970
- Includes information on at least 45 variables for each case, with more recent incidents including information on more than 120 variables
- Supervised by an advisory panel of 12 terrorism research experts
- Over 3,500,000 news articles and 25,000 news sources reviewed to collect incident data from 1998 to 2008 alone
- Available to Government representatives and interested researchers directly through their Online interface.

## III. AVAILABLE DATABASE VARIABLES

### A. *GTD ID (eventid) (Numeric)*

The incidents follow a 12-digit Event ID system.
The first 8 numbers – Recording date " yyyymmdd".
Next 2 numbers – always Zero Zero "00".
Last two numbers – case number for the given day (01,02 etc.) This will be 00 unless there is more than one case on the same date.
For example, an incident in the GTD occurring on 25 July 1993 would be numbered as "199307250001". An additional GTD case recorded for the same day would be "199307250002". The next GTD case recorded for that day would be "199307250003", etc.

To determine whether an incident is single, incidents occurring in both the same geographic and temporal point will be regarded as a single incident, but if either the *time* of occurrence of incidents or their *locations* are *discontinuous*, the events will be regarded as separate incidents.

Examples:

- *Four truck bombs explode nearly simultaneously in different parts of a major city.* This represents four incidents.

- *A bomb goes off, and while police are working on the scene the next day, they are attacked by terrorists with automatic weapons.* These are two separate incidents, as they were not continuous, given the time lag between the two events.

- *A group of militants shoot and kill five guards at a perimeter checkpoint of a petroleum refinery and then proceeds to set explosives and destroy the refinery.* This is one incident since it occurred in a single location (the petroleum refinery) and was one continuous event.

- *A group of hijackers diverts a plane to Senegal and, while at an airport in Senegal, shoots two Senegalese policemen.* This is one incident, since the hijacking was still in progress at the time of the shooting and hence the two events occurred at the same time in the same place.

If the information available for such complex events does not specify the time lag between or the exact locations of multiple terrorist activities, the event is a single incident.

## IV. INCIDENT DATE

### A. *ear (iyear) Numeric*

This field contains the year in which the incident occurred. In the case of incident(s) occurring over an extended period, the field will record the year when the incident was initiated. When the year of the incident is unknown, this will be recorded as “0”.

### B. *Month (imonth) Numeric*

This field contains the number of the month in which the incident occurred. In the case of incident(s) occurring over an extended period, the field will record the month when the incident was initiated. When the exact month of the incident is unknown, this will be recorded as “0”. For the cube this could form part of the Time dimension.

### C. *Day (iday) Numeric*

This field contains the numeric day of the month on which the incident occurred. In the case of incident(s) occurring over an extended period, the field will record the day when the incident was initiated.
When the exact day of the incident is unknown, the field is recorded as “0”.

### D. *Approximate Date (approxdate) Text*

Whenever the exact date of the incident is not known or remains unclear, this field is used to record the approximate date of the incident.

- If the day of the incident is not known, then the value for “Day” is “0”.
- For example, if an incident occurred in June 1978 and the exact day is not known, then the value for the “Day” field is “0” and the value for the “Approximate Date” field is “June 1978”.
- If the month is not known, then the value for the “Month” field is “0”.
- For example, if an incident occurred in the first half of 1978, and the values for the day and the month are not known, then the value for the “Day” and “Month” fields will both be “0” and the value for the “Approximate Date” field is “first half of 1978”.

### E. *Extended Incident? (extended) Categorical*

1 = "Yes" The duration of an incident extended more than 24 hours.
0 = "No" The duration of an incident extended less than 24 hours.

### F. *Date of Extended Incident Resolution (resolution)Date*

This field only applies if “Extended Incident?” is “Yes” and records the date in which the incident was resolved (hostages released by perpetrators; hostages killed; successful rescue, etc.)

It may be seen that variables categorised in this sub section Paras A to F can help form the Time dimension with the desired granularity.

## V. INCIDENT LOCATION

### A. *Country (country; country_txt) Categorical Variable*

This field identifies the country or location where the incident occurred. This includes non-independent states, dependencies, and territories, such as Northern Ireland and Corsica. If an incident occurs in an autonomous or geographically non-contiguous area, it is listed separately from the “home” country. However, separatist regions, such as Kashmir, Chechnya, South Ossetia, Transnistria, or Republic of Cabinda, are coded as part of the “home” country. West Bank and Gaza Strip have been coded separately from Israel. If an incident took place in a city located in the West Bank or Gaza Strip, it has been coded accordingly.
When an incident occurred in international waters or airspace, the country of departure is listed as the country of the incident. If the departure point is not identified, the incident is coded as “International.”
In cases where hostages were taken, the country where the incident began is recorded as the incident location, and a separate field captures the country where the incident was resolved or ended.

In the case where the country in which an incident occurred cannot be identified, it is coded as "Unknown The political circumstances of many countries have changed over time. In a number of cases, countries that represented the location of terrorist attacks no longer exist; examples include West Germany, the USSR and Yugoslavia. In these cases the country name for the year the event occurred is recorded. As an example, a 1989 attack in Bonn would be recorded as taking place in West Germany (FRG). An identical attack in 1991 would be recorded as taking place in Germany. The dates which apply as watersheds are as given below.
Eritrea – independence: 24 May 1993.
Germany – unification: 3 October 1990.
Breakup of Czechoslovakia
Czech Republic – independence: 1 January 1993; Slovakia – independence: 1 January 1993;
Breakup of USSR
Russian Federation – Independence: 24 August 1991; Armenia – Independence: 21 September 1991; Azerbaijan – Independence: 30 August 1991;
Belarus – independence: 25 August 1991;
Estonia – independence: 17 September 1991;
Georgia – independence: 9 April 1991;
Kazakhstan – independence: 16 December 1991;
Kyrgyzstan – independence: 31 August 1991;
Latvia – independence: 21 August 1991;
Lithuania – independence: 17 September 1991;
Moldova – independence: 27 August 1991;
Tajikistan – independence: 9 September 1991;
Turkmenistan – independence: 27 October 1991;
Ukraine – independence: 24 August 1991;
Uzbekistan – independence: 1 September 1991;
USSR terminates: 26 December 1991 – 5 January 1992.
Breakup of Yugoslavia:
Bosnia and Herzegovina – independence: 11 April 1992;
Croatia – independence: 25 June 1991;
Kosovo – independence: 17 February 2008:
Macedonia – independence: 8 September 1991;
Yugoslavia turns Serbia-Montenegro: 4 February 2003;
Montenegro – independence: 3 June 2006;
Serbia – independence: 3 June 2006
Slovenia – independence: 1 January 1992.

Country (Location) Codes (Note: These codes are also used for the target nationality fields)

4 = Afghanistan
5 = Albania
6 = Algeria
7 = Andorra
8 = Angola
10 = Antigua and Barbuda
11 = Argentina
12 = Armenia
14 = Australia
15 = Austria
16 = Azerbaijan
17 = Bahamas
18 = Bahrain
19 = Bangladesh
20 = Barbados
21 = Belgium
22 = Belize
23 = Benin
24 = Bermuda
25 = Bhutan
26 = Bolivia
28 = Bosnia-Herzegovina
29 = Botswana
30 = Brazil
31 = Brunei
32 = Bulgaria
33 = Burkina Faso
34 = Burundi
35 = Belarus
36 = Cambodia
37 = Cameroon
38 = Canada
40 = Cayman Islands
41 = Central African Republic
42 = Chad
43 = Chile
44 = China
45 = Colombia
46 = Comoros
47 = Congo (Brazzaville)
49 = Costa Rica
50 = Croatia
51 = Cuba
53 = Cyprus
54 = Czech Republic
55 = Denmark
56 = Djibouti
57 = Dominica
58 = Dominican Republic
59 = Ecuador
60 = Egypt
61 = El Salvador
62 = Equatorial Guinea
63 = Eritrea
64 = Estonia
65 = Ethiopia
66 = Falkland Islands
67 = Fiji
68 = Finland
69 = France
70 = French Guiana
71 = French Polynesia
72 = Gabon
73 = Gambia
74 = Georgia
75 = Germany
76 = Ghana
77 = Gibraltar

78 = Greece
79 = Greenland
80 = Grenada
81 = Guadeloupe
83 = Guatemala
84 = Guinea
85 = Guinea-Bissau
86 = Guyana
87 = Haiti
88 = Honduras
89 = Hong Kong
90 = Hungary
91 = Iceland
92 = India
93 = Indonesia
94 = Iran
95 = Iraq
96 = Ireland
97 = Israel
98 = Italy
99 = Ivory Coast
100 = Jamaica
101 = Japan
102 = Jordan
103 = Kazakhstan
104 = Kenya
106 = Kuwait
107 = Kyrgyzstan
108 = Laos
109 = Latvia
110 = Lebanon
111 = Lesotho
112 = Liberia
113 = Libya
115 = Lithuania
116 = Luxembourg
117 = Macau
118 = Macedonia
119 = Madagascar
120 = Malawi
121 = Malaysia
122 = Maldives
123 = Mali
124 = Malta
125 = Man, Isle of
127 = Martinique
128 = Mauritania
129 = Mauritius
130 = Mexico
132 = Moldova
134 = Mongolia
136 = Morocco
137 = Mozambique
138 = Myanmar
139 = Namibia
141 = Nepal
142 = Netherlands
143 = New Caledonia
144 = New Zealand
145 = Nicaragua
146 = Niger
147 = Nigeria
149 = North Korea
151 = Norway
152 = Oman
153 = Pakistan
155 = West Bank and Gaza Strip
156 = Panama
157 = Papua New Guinea
158 = Paraguay
159 = Peru
160 = Philippines
161 = Poland
162 = Portugal
163 = Puerto Rico
164 = Qatar
166 = Romania
167 = Russia
168 = Rwanda
173 = Saudi Arabia
174 = Senegal
175 = Serbia-Montenegro
176 = Seychelles
177 = Sierra Leone
178 = Singapore
179 = Slovak Republic
180 = Slovenia
181 = Solomon Islands
182 = Somalia
183 = South Africa
184 = South Korea
185 = Spain
186 = Sri Lanka
189 = St. Kitts and Nevis
195 = Sudan
196 = Suriname
197 = Swaziland
198 = Sweden
199 = Switzerland
200 = Syria
201 = Taiwan
202 = Tajikistan
203 = Tanzania
204 = Togo
205 = Thailand
207 = Trinidad and Tobago
208 = Tunisia
209 = Turkey
213 = Uganda
214 = Ukraine
215 = United Arab Emirates
216 = Great Britain
217 = United States
218 = Uruguay
219 = Uzbekistan
220 = Vanuatu

221 = Vatican City
222 = Venezuela
223 = Vietnam
225 = Virgin Islands (U.S.)
226 = Wallis and Futuna
227 = Samoa (Western Samoa)
228 = Yemen
229 = Congo (Kinshasa)
230 = Zambia
231 = Zimbabwe
233 = Northern Ireland
235 = Yugoslavia
236 = Czechoslovakia
238 = Corsica
296 = Kurdish
311 = Roma (Gypsy)
321 = Arab
334 = Asian
338 = African
347 = Timor-Leste
349 = Western Sahara
351 = Commonwealth of Independent States
359 = Soviet Union
362 = West Germany (FRG)
376 = Korea
377 = North Yemen
381 = Jewish
383 = Peru/U.S.
403 = Rhodesia
406 = South Yemen
422 = International
428 = South Vietnam
449 = Hindu
499 = East Germany (GDR)
512 = European
520 = Sinhalese
523 = Tuareg
529 = Middle Eastern
532 = New Hebrides
1003 = Kosovo

## B. *Region (region; region_txt) Categorical Variable*

This field identifies the region in which the incident occurred. The regions are divided into the following 13 categories:

1= North America ( Canada, Mexico, United States)
2= Central America & Caribbean (Antigua and Barbuda, Bahamas, Barbados, Belize, Bermuda, Cayman Islands, Costa Rica, Cuba, Dominica, Dominican Republic, El Salvador, Grenada, Guadeloupe, Guatemala, Haiti, Honduras, Jamaica, Martinique, Nicaragua, Panama, Puerto Rico, St. Kitts and Nevis, Trinidad and Tobago, Virgin Islands (U.S.))
3= South America (Argentina, Bolivia, Brazil, Chile, Colombia, Ecuador, Falkland Islands, French Guiana, Guyana, Paraguay, Peru, Suriname, Uruguay, Venezuela)
4= East Asia (China, Hong Kong, Japan, Macau, North Korea, South Korea, Taiwan)
5= Southeast Asia (Brunei, Cambodia, Indonesia, Laos, Malaysia, Myanmar, Philippines, Singapore, South Vietnam, Thailand, Timor-Leste, Vietnam)
6= South Asia (Afghanistan, Bangladesh, Bhutan, India, Maldives, Mauritius, Nepal, Pakistan, Seychelles, Sri Lanka)
7= Central Asia (Kazakhstan, Kyrgyzstan, Tajikistan, Uzbekistan)
8= Western Europe (Andorra, Austria, Belgium, Corsica, Denmark, East Germany (GDR), Finland, France, Germany, Gibraltar, Great Britain, Greece, Iceland, Ireland, Italy, Luxembourg, Malta, Man, Isle of, Netherlands, Northern Ireland, Norway, Portugal, Spain, Sweden, Switzerland, West Germany (FRG))
9= Eastern Europe (Albania, Bosnia-Herzegovina, Bulgaria, Croatia, Czech Republic, Czechoslovakia, Hungary, Kosovo, Macedonia, Moldova, Poland, Romania, Serbia-Montenegro, Slovak Republic, Slovenia, Yugoslavia)
10= Middle East & North Africa (Algeria, Bahrain, Cyprus, Egypt, Iran, Iraq, Israel, Jordan, Kuwait, Lebanon, Libya, Morocco, North Yemen, Qatar, Saudi Arabia, South Yemen, Syria, Tunisia, Turkey, United Arab Emirates, West Bank and Gaza Strip, Western Sahara, Yemen)
11= Sub-Saharan Africa (Angola, Benin, Botswana, Burkina Faso, Burundi, Cameroon, Central African Republic, Chad, Comoros, Congo (Brazzaville), Congo (Kinshasa), Djibouti, Equatorial Guinea, Eritrea, Ethiopia, Gabon, Gambia, Ghana, Guinea, Guinea-Bissau, Ivory Coast, Kenya, Lesotho, Liberia, Madagascar, Malawi, Mali, Mauritania, Mozambique, Namibia, Niger, Nigeria, Rhodesia, Rwanda, Senegal, Sierra Leone, Somalia, South Africa, Sudan, Swaziland, Tanzania, Togo, Uganda, Zambia, Zimbabwe)
12= Russia & the Newly Independent States (NIS) (Armenia, Azerbaijan, Belarus, Estonia, Georgia, Latvia, Lithuania, Russia, Soviet Union, Ukraine)
13= Australasia & Oceania (Australia, Fiji, French Polynesia, New Caledonia, New Hebrides, New Zealand, Papua New Guinea, Samoa (Western Samoa), Solomon Islands, Vanuatu, Wallis and Futuna)

## C. *Province / Administrative Region / U.S. State ((provstate) Text Variable)*

This variable records the name of the province, administrative region or U.S. State.

## D. *City (city) Text Variable*

This field contains the name of the city in which the incident occurred.

## E. *Vicinity (vicinity) Categorical Variable*

- 1 = "Yes" The incident occurred in the vicinity of the city in question.

- 0 = "No" The incident in the city itself.

### F. *Location Description (location) Text Variable*

This field is used to specify additional information about the location of the incident.

The above region, country, province fields could be used to form a a hierarchical Space Dimension starting at top level regions followed by country and then province. This level could further have the level of City but the data base has not covered this aspect in details as not all cities or provinces are covered by terror incidents.

## VI. INCIDENT INFORMATION

### A. *Incident Summary (summary) Text Variable*

A narrative summary of the incident, noting the “when, where, who, what, how, and why.” This field is available with incidents occurring after 1997.

### B. *Criteria Categorical Variables*

These variables record, as to which of the inclusion criteria (in addition to the necessary criteria) are met. This allows users to filter out those incidents whose inclusion was based on a criterion which they believe does not constitute terrorism proper.

***CRITERION 1: Political, Economic, Religious, or Social Goal (crit1)*** - The violent act must be aimed at attaining a political, economic, religious, or social goal. This criterion is not satisfied in those cases where the perpetrator(s) acted out of a pure profit motive or from an idiosyncratic personal motive unconnected with broader societal change.

1 = "Yes" The incident meets Criterion 1.
0 = "No" The incident does not meet Criterion 1.

***CRITERION 2: Intention to Coerce, Intimidate or Publicize to Larger Audience(s) (crit2)*** - To satisfy this criterion there must be evidence of an intention to coerce, intimidate, or convey some other message to a larger audience (or audiences) than the immediate victims. Such evidence can include (but is not limited to) the following: pre- or post-attack statements by the perpetrator(s), past behavior by the perpetrators, or the particular nature of the target, weapon, or attack type.
1 = "Yes" The incident meets Criterion 2.
0 = "No" The incident does not meet Criterion 2.

***CRITERION 3: Outside International Humanitarian Law (crit3)*** - The action must be outside the context of legitimate warfare activities, i.e. the act must be outside the parameters permitted by international humanitarian law (jus in bello) as reflected in the Additional Protocol to the Geneva Conventions of 12 August 1949 and elsewhere. Specifically, if an attack contravenes any of the following, this criterion is met:
Persons who are not, or are no longer, taking part in hostilities shall be respected, protected and treated humanely. They shall be given appropriate care, without any discrimination.
Captured combatants and other persons whose freedom has been restricted shall be treated humanely. They shall be protected against all acts of violence, in particular against torture. If put on trial, captured combatants shall enjoy the fundamental guarantees of a regular judicial procedure.
The right of parties to an armed conflict to choose methods or means of warfare is not unlimited. No superfluous injury or unnecessary suffering shall be inflicted.
In order to spare the civilian population, armed forces shall at all times distinguish between the civilian population and civilian objects on the one hand, and military objectives on the other. Neither the civilian population as such nor individual civilians or civilian objects shall be the targets of military attacks.

1 = "Yes" The incident meets Criterion 3.
0 = "No" The incident does not meet Criterion 3.

### C. *Doubt Terrorism Proper? (doubtterr) Categorical Variable*

In certain cases there may be some uncertainty whether an incident meets all of the criteria for inclusion. In these ambiguous cases, where there is a strong possibility, but not certainty, that an incident represents an act of terrorism, the incident is included in GTD and is coded as “Yes” for this variable.

- 1 = "Yes" There is doubt as to whether the incident is an act of terrorism.
- 0 = "No" There is essentially no doubt as to whether the incident is an act of terrorism.

This field is presently only available with incidents occurring after 1997. Incidents occurring before 1998 are coded as “-9” for this variable.

### D. *Alternative Designation (alternative; alternative_txt) Categorical Variable*

This variable applies to only those cases coded as “Yes” for “Doubt Terrorism Proper?” (above). This variable identifies the most likely categorization of the incident other than terrorism.

- 1= Insurgency/Guerilla Action
- 2= Purely Criminal Act
- 3= Mass Murder
- 4= Internecine Conflict Action

This field is presently only available with incidents occurring after 1997.

### E. *Part of Multiple Incident (multiple) Categorical Variable*

In those cases where several attacks are connected, but where the various actions do not constitute a single incident (either the time of occurrence of incidents or their locations are discontinuous), then “Yes” is selected to denote that the particular attack was part of a “multiple” incident.

- 1 = "Yes" The attack is part of a multiple incident.
- 0 = "No" The attack is not part of a multiple incident.

### F. *Situation of Multi-Party Conflict (conflict) Categorical Variable*

When there are multiple groups in conflict, and some of the groups might be committing terrorist acts, it is often difficult to attribute responsibility or to unequivocally discern various non-state actors. In this case, “Yes” is selected.

- 1 = "Yes" The incident took place in the context of a multi-party conflict.
- 0 = "No" The incident did not take place in the context of a multi-party conflict.

## VII. ATTACK INFORMATION

### A. *Successful Attack (success) Categorical Variable*

Success of a terrorist strike is defined according to the tangible effects of the attack. For example, in a typical successful bombing, the bomb detonates and destroys property and/or kills individuals, whereas an unsuccessful bombing is one in which the bomb is discovered and defused or detonates early and kills the perpetrators. Success is not judged in terms of the larger goals of the perpetrators. For example, a bomb that exploded in a building would be counted as a success even if it did not, for example, succeed in bringing the building down or inducing government repression.

1 = "Yes" The incident was successful.
0 = "No" The incident was not successful.

### B. *Suicide Attack (suicide) Categorical Variable*

This variable is coded “Yes” in those cases where there is evidence that the perpetrator did not intend to escape from the attack alive.

1 = "Yes" The incident was a suicide attack.
0 = "No" The incident was not a suicide attack.

### C. *Attack Type (attacktype1; attacktype1_txt) Categorical Variable*

Up to three attack types are recorded for each incident. This field captures the general method of attack and often reflects the broad class of tactics used. It consists of the following nine categories, which are defined below:

*1= Assassination* An act whose primary objective is to kill one or more specific, prominent individuals. Usually carried out on persons of some note, such as high-ranking military officers, government officials, celebrities, etc. Not to include attacks on non-specific members of a targeted group. The killing of a police officer would be an armed assault unless there is reason to believe the attackers singled out a particularly prominent officer for assassination.

*2= Armed Assault* An attack whose primary objective is to cause physical harm or death directly to human beings by any means other than an explosive.

*3= Bombing/Explosion* An attack where the primary effects are caused by an energetically unstable material undergoing rapid decomposition (either deflagration or detonation) and releasing a pressure wave that causes physical damage to the surrounding environment. Can include either high or low explosives but does not include a nuclear explosive device that releases energy from fission and/or fusion, or an incendiary device where decomposition takes place at a much slower rate.

*4=Hijacking* An act whose primary objective is to take control of a vehicle such as an aircraft, boat, bus, etc. for the purpose of diverting it to an unprogrammed destination, obtain payment of a ransom, force the release of prisoners, or some other political objective. Hijackings are distinct from Hostage Taking because the target is a vehicle, regardless of whether there are people/passengers in the vehicle.

*5=Hostage Taking (Barricade Incident)* An act whose primary objective is to obtain political or other concessions in return for the release of prisoners (hostages). Such attacks are distinguished from kidnapping since the incident occurs and usually plays out at the target location with little or no intention to hold the hostages for an extended period in a separate clandestine location.

*6=Hostage Taking (Kidnapping)* As for Barricade Incident above, but distinguished by the intention to move and hold the hostages in a clandestine location. Usually in kidnappings the victims are selected beforehand.

*7=Facility / Infrastructure Attack* An act, excluding the use of an explosive, whose primary objective is to cause damage to a non-human target, such as a building, monument, train, pipeline, etc. Such attacks consist of actions primarily aimed at damaging property, or at causing a diminution in the functioning of a useful system (mass disruption) yet not causing direct harm to people. Such attacks include arson, cyber attacks, and various forms of sabotage. Can include acts that intend to cause harm to people as a result of the harm done to objects (e.g., blowing up a dam so that the ensuing flood will kill residents downstream). Can include acts which aim to harm an installation, yet also cause harm to people incidentally.

*8=Unarmed Assault* An attack whose primary objective is to cause physical harm or death directly to human beings by any means other than explosive, firearm, incendiary, or sharp instrument (knife, etc.).

*9=Unknown* The attack type cannot be determined from the available information.

D. *Second Attack Type (attacktype2; attacktype2_txt) Categorical Variable* – **Coding is same as above.**

E. *Third Attack Type (attacktype3; attacktype3_txt) Categorical Variable-* – **Coding is same as above.**

## VIII. TARGET INFORMATION

Information on up to three targets is recorded for each incident. The target information fields coded for each of the three targets include target type, target entity, name of entity, specific target, and nationality of the target.

A. *A. Target Type (targtype1; targtype1_txt) Categorical Variable*

The target type field captures the general type of target. It consists of the following 22 categories, which are defined as under:

***1=Business*** - Businesses are defined as individuals or organizations engaged in commercial or mercantile activity as a means of livelihood. Any attack on a business or private citizens patronizing a business such as a restaurant, gas station, music store, bar, café, etc. This includes attacks carried out against corporate offices or employees of firms like mining companies, or oil corporations. Furthermore, includes attacks conducted on business people or corporate officers. Included in this value as well are hospitals and chambers of commerce and cooperatives. It does not include attacks carried out in public or quasi-public areas such as "business district or commercial area", (these attacks are captured under "Private Citizens and Property", see below.)

***2=Government (General)*** - Any attack on a government building; government member, former members, including members of political parties, their convoys, or events sponsored by political parties; political movements; or a government sponsored institution where the attack is expressly carried out to harm the government. This value includes attacks on judges, public attorneys (e.g., prosecutors), courts and court systems, politicians, royalty, head of state, government employees (unless police or military), election-related attacks, intelligence agencies and spies.

***3=Police*** – This value includes attacks on members of the police force or police installations; this includes police boxes, patrols, Headquarters, academies, cars, checkpoints, etc. This includes attacks against jails or prison facilities, or jail or prison staff or guards. Also includes attacks against private security guards and security forces.

***4= Military*** - Includes attacks against army units, patrols, barracks, and convoys, jeeps, etc. Also includes attacks on recruiting sites, and soldiers engaged in internal policing functions such as at checkpoints and in anti-narcotics activities. It excludes attacks against militia and guerrillas, these types of attacks are coded as "Terrorist" see below.

***5=Abortion Related*** - Attacks on abortion clinics, employees, patrons, or security personnel stationed at clinics.

***6=Airports & Airlines*** – An attack that was carried out either against an airplane or against an airport. Attacks against airline employees while on board are also included in this value. It includes attacks conducted against airport business offices and executives. Attacks where airplanes were used to carry out the attack (such as three of the four 9/11 attacks) are not included.

***7=Government (Diplomatic)*** - Attacks carried out against foreign missions, including embassies, consulates, etc. This value includes cultural centers that have diplomatic functions, and attacks against diplomatic staff and their families and property.

***8=Educational Institution*** - Attacks against schools, teachers, or guards protecting school sites. Includes attacks against university professors, teaching staff and school buses. Moreover, includes attacks against religious schools in this value. As noted below in the "Private Citizens and Property" value, the database has several attacks against students. If attacks involving students are not expressly against a school, university or other educational institution or are carried out in an educational setting, they are coded as private citizens and property. This excludes attacks against military schools (attacks on military schools are coded as "Military,").

***9=Food or Water Supply*** - Attacks on food or water supplies or reserves are included in this value.

***10=Journalists & Media*** - Includes, attacks on reporters, news assistants, photographers, publishers, as well as attacks on media headquarters and offices. Attacks on transmission facilities such as antennae or transmission towers are included in this value (while attacks on broadcast infrastructure are coded as "Telecommunications,").

***11=Maritime (Includes Ports and Maritime Facilities)*** - Implies civilian maritime. Includes attacks against fishing ships, oil tankers, ferries, yachts, etc. (Attacks on fishermen are coded as "Private Citizens and Property," see below).

- ***12=NGO*** - Includes attacks on offices and employees of non-governmental organizations (NGOs). NGOs here are defined as primarily large multinational non-governmental organizations. These include the Red Cross and Doctors without Borders. Peacekeepers also belong to this value. This does not include labor unions, social clubs, student groups, and other non-NGO (such cases are coded as "Other".).
- ***13=Other*** - This value includes acts of terrorism committed against targets which do not fit into other categories.
- ***14=Private Citizens & Property*** -This value includes attacks on individuals, the public in general or attacks in public areas including markets, commercial streets, busy intersections and pedestrian malls. This also includes ambiguous cases where the target was a named individual, or where the target/victim of an attack could be identified by name, age, occupation, gender or nationality. This value also includes ceremonial events, such as weddings and funerals. The database contains a number of attacks against students. If these attacks are not expressly against a school, university or other educational institution or are not carried out in an educational setting, these attacks are coded using this value. Also, includes incidents involving political supporters as private citizens and property, provided that these supporters are not part of a government-sponsored event. Finally, this value includes police informers. This does not include attacks causing civilian casualties in businesses such as restaurants, cafes or movie theaters (these categories are coded as "Business" see above).

- ***15= Religious Figures/ Institutions*** - This value includes attacks on religious leaders, (Imams, priests, bishops, etc.), religious institutions (mosques, churches), religious places or objects (shrines, relics, etc.). This value also includes attacks on organizations that are affiliated with religious entities that are not NGOs, businesses or schools. Attacks on religious pilgrims are considered "Private Citizens and Property;" attacks on missionaries are considered religious figures.
- ***16=Telecommunication*** - This includes attacks on facilities and infrastructure for the transmission of information. More specifically this value includes things like cell phone towers, telephone booths, television transmitters, radio, and microwave towers.
- ***17=Terrorists*** - Terrorists or members of identified terrorist groups are included in this value. Membership is broadly defined and includes informants for terrorist groups, but excludes former terrorists. This value also includes cases involving the targeting of militias and guerillas.
- ***18=Tourists*** - This value includes the targeting of tour buses, tourists, or "tours." Tourists are persons who travel primarily for the purposes of leisure or amusement. Government tourist offices are included in this value. The attack must clearly target tourists, not just an assault on a business or transportation system used by tourists.
- ***19=Transportation (Other than Aviation)*** - Attacks on public transportation systems are included in this value. This can include efforts to assault public buses, minibuses, trains, metro/subways, highways (if the highway itself is the target of the attack), bridges, roads, etc. The database contains a number of attacks on generic terms such as "cars" or "vehicles." These attacks are assumed to be against "Private Citizens and Property" unless shown to be against public transportation systems. In this regard, buses are assumed to be public transportation unless otherwise noted.
- ***20=Unknown*** - The target type cannot be determined from the available information.
- ***21=Utilities*** - This value pertains to facilities for the transmission or generation of energy. For example, power lines, oil pipelines, electrical transformers, high tension lines, gas and electric substations, are all included in this value. This value also includes lampposts or street lights. Attacks on officers, employees or facilities of utility companies excluding the type of faculties above are coded as business.
- ***22=Violent Political Parties*** -This value pertains to entities that are both political parties (and thus, coded as "government" in this coding scheme) and terrorists. It is operationally defined as groups that engage in electoral politics and appear as "Perpetrators" in the database.

### B. *Target Entity (entity1; entity1_txt) Categorical Variable*

The entity field refers to the type of organization or interest group represented by the specific target attacked, and provides an alternate categorization to "Target Type" above.

1 = Diplomat
2 = Police/Military
3 = Other
4 = Unknown
5 = Government
6 = Political Party
7 = Media
8 = Business
9 = Transportation
10= Utilities
11 = Foreign Business
12 = Domestic Business
13 = Transportation
14 = Utilities
15 = Media
16 = Diplomat
17 = Government
18 = International
19 = Other
20 = Police/Military
21 = Political Party
22 = Unknown
23 = Religious Figures/Institutions
24 = Indiscriminate Civilians/Non-Combatants
25 = Religious Figures/Institutions
26 = Indiscriminate Civilians/Non-Combatants

### C. *Name of Entity (corp1) Text Variable*

This is the name of the corporate entity or government agency that was targeted. If no specific entity was targeted, this field is left blank. If the element targeted is unspecified, "Unknown" is listed.

*D. Specific Target (target1) Text Variable*

This is the specific person, building, installation, etc., that was targeted and is a part of the entity named above. (For example, if the U.S. Embassy in Country X was attacked the "Name of Entity" would be "U.S. Department of State" and the "Specific Target" would be "U.S. Embassy in Country X.") However, if the target includes multiple victims (e.g., in a kidnapping or assassination), only the first victim's name is recorded in this field, with remaining names recorded in the "Additional Notes" field.

*E. Nationality of Target (natlty1; natlty1_txt) Categorical Variable*

This is the nationality of the target that was attacked, and is not necessarily the same as the country in which the incident occurred, although in most cases it is. For hijacking incidents, the nationality of the plane is recorded and not that of the passengers. Numeric nationality codes are same as the country codes .

*F. Second Target Type (targtype2; targtype2_txt) Categorical Variable* – Same as targtype1 above.

*G. Second Target Entity (entity2; entity2_txt) Categorical Variable – Same as entity1 above.*

*H. Name of Second Entity (corp2) Text Variable*

Same as "Name of Entity" field.

*I. Second Specific Target (target2) Text Variable*

Conventions follow "Specific Target" field.

## IX. NATIONALITY OF SECOND TARGET (NATLTY2; NATLTY2_TXT) CATEGORICAL VARIABLE

Conventions follow "Nationality of Target" field. For numeric nationality codes, as per the country codes in section V above.

*A. Third Target Type (targtype3; targtype3_txt) Categorical Variable*

Conventions follow "Target Entity" field.

*B. Name of Third Entity (corp3) Text Variable*

Conventions follow "Name of Entity" field.

*C. Third Specific Target (target3) Text Variable*

Conventions follow "Specific Target" field.

*D. Nationality of Third Target (natlty3; natlty3_txt) Categorical Variable*

Conventions follow "Nationality of Target" field. For numeric nationality codes, please see the country codes in section III-A.

## X. PERPETRATOR INFORMATION

Information on up to three perpetrators is recorded for each incident. This includes the perpetrator group name and the perpetrator group sub-name, in addition to the specific motive of the attack and a record of whether or not the attribution of responsibility is unconfirmed.

*A. Perpetrator Group Name (gname) Text Variable*

This field contains the name of the group that carried out the attack. In order to ensure consistency in the usage of group names for the database, the GTD database uses a standardized list of group names that have been established by project staff to serve as a reference for all subsequent entries.

*B. Perpetrator Sub-Group Namen(gsubname) Text Variable*

This field contains any additional qualifiers or details about the name of the group that carried out the attack. This includes but is not limited to the name of the specific faction when available.

*C. Second Perpetrator Group Name (gname2) Text Variable*

This field is used to record the name of the second perpetrator when responsibility for the attack is attributed to more than one perpetrator. Conventions follow "Perpetrator Group" field.

*D. Second Perpetrator Sub-Group Name (gsubname2) Text Variable*

This field is used to record additional qualifiers or details about the second perpetrator group name when responsibility for the attack is attributed to more than one perpetrator. Conventions follow "Perpetrator Sub-Group Name" field.

*E. Third Perpetrator Group Name (gname3) Text Variable*

This field is used to record the name of the third perpetrator when responsibility for the attack is attributed to more than two perpetrators. Conventions follow "Perpetrator Group" field.

*F. Third Perpetrator Sub-Group Name (gsubname3) Text Variable*

This field is used to record additional qualifiers of details about the third perpetrator group name when responsibility for the attack is attributed to more than two perpetrators. Conventions follow "Perpetrator Sub-Group Name" field.

*G. Specific Motive (motive) Text Variable*

When reports explicitly mention a specific motive for the attack, this motive is recorded in the "Specific Motive" field.

*H. Perpetrator Group(s) Suspected/Unconfirmed? (guncertain) Categorical Variable*

- "Yes" is used in circumstances where a government official is reported to be expressing a suspicion, or educated guess or other unconfirmed / speculative position regarding the identity of the terrorist group mounting the attack. Cases where credible, non-government analysts identify probable perpetrators receive a "No" in this field.
- Cases where a terrorist group claims responsibility for the attack are recorded as "No" unless the source specifically notes that authorities doubt the veracity of the claim.
- Cases where a government official expresses a definite position on the perpetrator based on intelligence or other information are recorded as "No".
- 1 = "Yes" The perpetrator attribution(s) for the incident are unconfirmed.
- 0 = "No" The perpetrator attribution(s) for the incident are not unconfirmed.

## XI. PERPETRATOR STATISTICS

*A. Number of Perpetrators (nperps)Numeric Variable*

This field indicates the total number of terrorists participating in the incident. (In the instance of multiple perpetrator groups participating in one case, the total number of perpetrators, across groups, is recorded). There are often discrepancies in information on this value.

Where several independent credible sources1 report different numbers of attackers, the value of this variable reflects the number given by the majority of sources, unless there is reason to do otherwise. Where there is no majority figure among independent sources, the database records the lowest preffered perpetrator figure, unless there is clear reason to do otherwise. In cases where the number of perpetrators is stated vaguely, for example "…at least 11 attackers", then the lowest possible number is recorded, in this example, "11." "-99" or "Unknown" appears when the number of perpetrators is not reported.

*B. Number of Perpetrators Captured (nperpcap) Numeric Variable*

This field records the number of perpetrators taken into custody.

- "-99" or "Unknown" appears when there is evidence of captured, but the number is not reported.
- Divergent reports on the number of perpetrators captured are dealt with in same manner used for the Number of Perpetrators variable described above.

## XII. PERPETRATOR CLAIM OF RESPONSIBILITY

*A. A. Claim of Responsibility?(claimed) Categorical Variable*

This field is used to indicate whether a group or person(s) claimed responsibility for the attack. If marked "Yes", it indicates that a person or a group did in fact claim responsibility. When there are multiple perpetrator groups involved, this field refers to the First Perpetrator Group (separate fields for the Second and Third groups follow below).

- 1 = "Yes" A group or person claimed responsibility for the attack.
- 0 = "No" No claim of responsibility was made.
- -9 = "Unknown" It is unknown whether or not a claim of responsibility was made.

*B. Mode for Claim of Responsibility (claimmode; claimmode_txt) Categorical Variable*

This records one of 10 modes used by claimants to claim responsibility and might be useful to verify authenticity, track trends in behavior, etc. If greater detail exists (for instance, a particularly novel or strange mode is used) this information is captured in the "Additional Notes" field.

Mode Values:
1 = Letter
2 = Call (post-incident)
3 = Call (pre-incident)
4 = E-mail
5 = Note left at scene
6 = Video
7 = Posted to website, blog, etc.
8 = Personal claim
9 = Other
10 = Unknown

*C. Claim Confirmed? (claimconf) Categorical Variable*

"Yes" or "No", indicate whether or not the claim is confirmed. "Unknown" appears if this information is not available.

*D. Second Group Claim of Responsibility? (claim2) Categorical Variable*

1 = "Yes" A group or person claimed responsibility for the attack.
0 = "No" No claim of responsibility was made.
-9 = "Unknown" It is unknown whether or not a claim of responsibility was made.
Conventions follow "Claim of Responsibility" field.

*E. Mode for Second Group Claim of Responsibility*

*(claimmode2; claimmode2_txt) Categorical Variable*

Conventions follow "Mode for Claim of Responsibility" field.

*F. Second Group Claim of Responsibility Confirmed? (claimconf2) Categorical Variable*

Conventions follow "Claim of Responsibility Confirmed?" field.

*G. Third Group Claim of Responsibility? (claim3) Categorical Variable*

Conventions follow "Claim of Responsibility" field.

*H. Mode for Third Group Claim of Responsibility (claimmode3; claimmode3_txt) Categorical Variable*

Conventions follow "Mode for Claim of Responsibility" field.

*I. Third Group Claim of Responsibility Confirmed? (claimconf3) Categorical Variable*

Conventions follow "Claim of Responsibility Confirmed?" field.

*J. Competing Claims of Responsibility? (compclaim) Categorical Variable*

This field is used to indicate whether more than one group claimed separate responsibility for the attack. If marked "Yes", it indicates that the groups entered in conjunction with the case each claimed responsibility for the attack (i.e., they did not work together, but each independently tried to claim credit for the attack).

- 1 = "Yes" There are competing claims of responsibility for the attack.
- 0 = "No" There are not competing claims of responsibility for the attack.
- -9 = "Unknown" It is unknown whether or not the claim of responsibility is confirmed.

## XIII. WEAPON INFORMATION

Information on up to four types and sub-types of the weapons used in an attack are recorded for each case, in addition to any information on specific weapon details reported.

*A. Weapon Type (weaptype1; weaptype1_txt) Categorical Variable*

This field records the general type of weapon used in the incident. It consists of the following 13 categories:

1 = Biological
2 = Chemical
3 = Radiological
4 = Nuclear
5 = Firearms
6 = Explosives/Bombs/Dynamite
7 = Fake Weapons
8 = Incendiary
9 = Melee
10 = Vehicle (not to include vehicle-borne explosives, i.e., car or truck bombs)
11 = Sabotage Equipment
12 = Other
13 = Unknown

*B. Weapon Sub-type (weapsubtype1; weapsubtype1_txt) Categorical Variable*

This field records a more specific value for most of the Weapon Types identified immediately above.
Values for Weapon Type and corresponding Sub-type

- Biological [no corresponding weapon sub-types]
- Chemical
  1 = Poisoning
- Radiological [no corresponding weapon sub-types]
- Nuclear [no corresponding weapon sub-types]
- Firearms
  2 = Automatic Weapon
  3 = Handgun
  4 = Rifle/Shotgun (non-automatic)
  5 = Unknown Gun Type
  6 = Other Gun Type
- Explosives/Bombs/Dynamite
  7 = Grenade
  8 = Land Mine
  9 = Letter Bomb
  10 = Pressure Trigger
  11 = Projectile (rockets, mortars, RPGs, etc.)
  12 = Remote Trigger
  13 = Suicide (carried bodily by human being)
  14 = Time Fuse
  15 = Vehicle
  16 = Unknown Explosive Type
  17 = Other Explosive Type
- Fake Weapons [no corresponding weapon sub-types]
- Incendiary
- Melee
  18 = Arson/Fire
  19 = Flame Thrower
  20 = Gasoline or Alcohol
  21 = Blunt Object
  22 = Hands, Feet, Fists
  23 = Knife
  24 = Rope or Other Strangling Device
  25 = Sharp Object Other Than Knife
  26 = Suffocation
- Vehicle (not to include vehicle-borne explosives, i.e., car or truck bombs) [no corresponding weapon sub-types]
- Sabotage Equipment [no corresponding weapon sub-types]
- Other [no corresponding weapon sub-types]
- Unknown [no corresponding weapon sub-types]

*C. Second Weapon Type (weaptype2; weaptype2_txt) Categorical Variable*

Conventions follow “Weapon Type” field.

*D. Second Weapon Sub-Type (weapsubtype2; weapsubtype2_txt) Categorical Variable*

Conventions follow “Weapon Sub-Type” field.

*E. Third Weapon Type (weaptype3; weaptype3_txt) Categorical Variable*

Conventions follow “Weapon Type” field.

*F. Third Weapon Sub-Type (weapsubtype3; weapsubtype3_txt) Categorical Variable*

Conventions follow “Weapon Sub-Type” field.

*G. Fourth Weapon Type (weaptype4; weaptype4_txt) Categorical Variable*

Conventions follow “Weapon Type” field.

*H. Fourth Weapon Sub-Type (weapsubtype4; weapsubtype4_txt) Categorical Variable*

Conventions follow “Weapon Sub-Type” field.

*I. Weapon Details (weapdetail) Text Variable*

This field notes any pertinent information on the type of weapon(s) used in the incident. Such notes could include the novel use or means of concealing a weapon, specific weapon models, interesting details of the weapons' origins, etc.

## XIV. CASUALTY INFORMATION

If several cases are linked together, the open-source reports sometimes list the number of casualties cumulatively. In such cases the preservation of statistical accuracy is preserved by the GTD by evenly distributing casualties across the linked incidents.

*A. Total Number of Fatalities (nkill) Numeric Variable*

- This field stores the number of total confirmed fatalities for the incident. The number includes all victims and attackers who died as a direct result of the incident.
- Where there is evidence of fatalities, but the number is not reported, “-99”or “Unknown” is the value given to this field.
- Where several independent sources report different numbers of casualties, the database will usually reflect the number given by the most recent source, unless there is reason to do otherwise. Where there are several “most recent” sources published around the same time, then the majority figure will be used. Where there is no majority figure among independent sources, the database will record the lowest proffered fatality figure, unless there is clear reason to do otherwise.

*B. Number of U.S. Fatalities (nkillus) Numeric Variable*

Limited to only U.S. fatalities, this field follows the conventions of “Total Number of Fatalities” above.

*C. Number of Perpetrator Fatalities (nkillter)Numeric Variable*

Limited to only perpetrator fatalities, this field follows the conventions of “Total Number of Fatalities” field.

*D. Total Number of Injured (nwound) Numeric Variable*

This field records the number of confirmed non-fatal injuries. Conventions follow the “Total Number of Fatalities” field.

*E. Number of U.S. Injured (nwoundus) Numeric Variable*

Conventions follow the “Number of U.S. Fatalities” field.

*F. Number of Perpetrators Injured (nwoundte) Numeric Variable*

Conventions follow the “Number of Perpetrator Fatalities” field.

## XV. CONSEQUENCES

*A. Property Damage? (property) Categorical Variable*

“Yes” appears if there is evidence of property damage during the incident.

- 1 = "Yes" The incident resulted in property damage.
- 0 = "No" The incident did not result in property damage.
- -9 = "Unknown" It is unknown whether or not the incident resulted in property damage

*B. Extent of Property Damage (propextent; propextent_txt) Categorical Variable*

If “Property Damage?” is “Yes” then one of four categories describe the extent of the property damage:
1 = Catastrophic (likely > $1 billion)
2 = Major (likely > $1 million but < $1 billion)
3 = Minor (likely < $1 million)
4 = Unknown

*C. Value of Property Damage (in U.S. $) (propvalue) Numeric Variable*

If “Property Damage?” is “Yes” then the exact U.S. dollar amount (at the time of the incident) of total damages is listed. If no dollar figure is reported, the field is blank. That is, a blank field here does not indicate that there was no property damage but, rather, that no precise estimate of the value was available. The value of damages only includes direct economic effects of the incident (i.e. cost of buildings, etc.) and not indirect economic costs (longer term effects on the company, industry, tourism, etc.). Protocols for recording inconsistent numbers, etc., listed above are followed (see, for example, “Number of Perpetrators”).

*D. Property Damage Comments (propcomment) Text Variable*

If “Property Damage?” is “Yes” then non-monetary or imprecise measures of damage may be described in this field.

## XVI. HOSTAGE / KIDNAPPING ADDITIONAL INFORMATION

*A. Hostages or Kidnapping Victims? (ishostkid) Categorical Variable*

This field records whether or not the victims were taken hostage or kidnapped.

- 1 = "Yes" The victims were taken hostage or kidnapped.
- 0 = "No" The victims were not taken hostage or kidnapped.
- -9 = "Unknown" It is unknown whether or not the victims were taken hostage or kidnapped.

*B. Total Number of Hostages/ Kidnapping Victims (nhostkid) Numeric Variable*

This field records the total number of hostages or kidnapping victims. As with the number of perpetrators, where several independent sources report different numbers of hostages, the GTD reflects the number given by the majority of sources, unless there is reason to do otherwise. Where there is no majority figure among independent sources, the database will record the lowest proffered hostage figure, unless there is clear reason to do otherwise. In cases where the number of hostages or kidnapping victims is stated vaguely, for example, “…at least 11 hostages”, then the lowest possible number will be recorded, in this example “11.” If the number of hostages is unknown or unidentified, this field records “-99” or “Unknown.”

*C. Number of U.S. Hostages/ Kidnapping Victims (nhostkidus) Numeric Variable*

Conventions follow the “Total Number of Hostages/ Kidnapping Victims” field, but only include U.S. hostage/kidnapping victims.

*D. Hours of Kidnapping / Hostage Incident (nhours) Numeric Variable*

- If the “Attack Type” is “Hostage Taking (Kidnapping),” “Hostage Taking (Barricade Incident),” or “Hijacking” then the duration of the incident is recorded either in this field or in the next field.
- If the incident lasted for less than 24 hours, this field records the number of hours.
- If the incident lasts for more than 24 hours (i.e., at least one day), then the number of days is recorded in the next field.

*E. Days of Kidnapping / Hostage Incident (ndays) Numeric Variable*

If the “Attack Type” is “Hostage Taking (Kidnapping),” “Hostage Taking (Barricade Incident),” or “Hijacking” and if the duration of the kidnapping / hostage incident last for more than 24 hours, this field records the duration of the incident in days. If information on hours and days is provided, the figure is rounded to the nearest day.

*F. Country That Kidnappers/Hijackers Diverted To (divert) Text Variable*

If the “Attack Type” is “Hostage Taking (Kidnapping)” or “Hijacking” then this field lists the country that the hijackers diverted the vehicle to. If the hijackers did not divert the vehicle to another country, this field is blank.

*G. Country of Kidnapping/Hijacking Resolution (kidhijcountry) Text Variable*

If the “Attack Type” is “Hostage Taking (Kidnapping)” or “Hijacking” then this field lists the country in which the incident was resolved or ended. If the incident was not resolved in another country, this field is blank.

*H. Ransom Demanded? (ransom) Categorical Variable*

“Yes” is recorded if the incident involved the demand of some form of ransom.

- 1 = "Yes" The incident involved a demand of ransom.
- 0 = "No" The incident did not involve a demand of ransom.
- -9 = "Unknown" It is unknown whether or not the incident involved a demand of ransom.

*I. Total Ransom Amount Demanded (ransomamt) Numeric Variable*

If a ransom was demanded then the amount of ransom demanded is listed in U.S. dollars. If a ransom was demanded but the monetary figure was unknown then this field is recorded with “-99” or “Unknown.”

*J. Ransom Amount Demanded from U.S. Sources (ransomamtus) Numeric Variable*

If a ransom was demanded from U.S. sources then this figure is listed in U.S. dollars. If a ransom was demanded from U.S. sources but the monetary figure was unknown then this field is recorded with “-99” or “Unknown.”

*K. Total Ransom Amount Paid (ransompaid) Numeric Variable*

If a ransom amount was paid then this figure is listed in U.S. dollars. If a ransom was paid but the monetary figure was unspecified then this field is recorded with "-99" or "Unknown."

*L. Ransom Amount Paid By U.S. Sources (ransompaidus) Numeric Variable*

If a ransom amount was paid by U.S. sources then this figure is listed in U.S. dollars. If a ransom was paid by U.S. sources but the monetary figure was unspecified then this field is recorded with "-99" or "Unknown."

*M. Ransom Notes (ransomnote) Text Variable*

If a ransom was demanded this field may be used to record any specific comments relating to the ransom not captured in other fields.

*N. Kidnapping/Hostage Outcome (hostkidoutcome; hostkidoutcome_txt) Categorical Variable*

If the "Attack Type" is "Hostage Taking (Kidnapping)" then this field applies. The seven values for this field are:

1 = Attempted Rescue
2 = Hostage(s) released by perpetrators
3 = Hostage(s) escaped (not during rescue attempt)
4 = Hostage(s) killed (not during rescue attempt)
5 = Successful Rescue
6 = Combination
7 = Unknown

If the hostages suffered a variety of the above fates, "Combination" is selected. Further details about the fate of hostages may be recorded in the "Additional Notes" field.

*O. Number Released/Escaped/Rescued (nreleased) Numeric Variable*

If the "Attack Type" is "Hostage Taking (Kidnapping)" then this field will apply. This field records the number of hostages who survived the incident. All previous protocols for recording numbers apply, including using "-99" for "Unknown."

As with the total number of kidnapping victims, where several independent sources report different numbers of hostages, the database will reflect the number given by the majority of sources, unless there is reason to do otherwise. Where there is no majority figure among independent sources, the database will record the lowest proffered hostage released/escaped/rescued figure, unless there is clear reason to do otherwise. In cases where the number of hostages released/escaped/rescued is stated vaguely, for example "…at least 11 hostages were released", then the lowest possible number will be recorded, in this example "11". If the number of hostages released/escaped/rescued is unknown or unidentified, this is recorded as "Unknown".

## XVII. ADDITIONAL INFORMATION

*A. (addnotes) Text Variable*

This field is used to capture the following information:

- Additional information that could not be captured in any of the above fields, such as details about hostage conditions or additional countries hijacked vehicles were diverted to.
- Supplemental important information not specific to the particular attack, such as multiple attacks in the same area or by the same perpetrator.
- Uncertainties about the data (such as differing reports of casualty numbers or perpetrators responsible).
- Unusual factors, such as a shift in tactics, the reappearance of an organization, the emergence of a new organization, an attack carried out on a historical date, or an escalation of a violent campaign.
- The fate (legal, health, or otherwise) of either victims or perpetrators where this is mentioned in GTD source documents.

*B. In addition, the instructions for several fields listed above have specific indications for placing additional information in this "Additional Notes" field, as needed:*

- Specific Target - If the Target is multiple victims (e.g., in a kidnapping or assassination), only the first name is recorded in the "Specific Target" field, with remaining names recorded in the "Additional Notes" field.
- Perpetrator Individual(s)' Name(s) - Names of individuals identified as planners, bomb-makers, etc., who are indirectly involved in an attack, may recorded in the "Additional Notes" field.
- Mode for Claim of Responsibility - If greater detail is needed than provided for the "Mode for Claim of Responsibility" field (for instance, a particularly novel or strange mode is used) this information may be captured in the "Additional Notes" field.
- Kidnapping/Hostage Outcome - If greater detail is available than the Kidnapping/Hostage Outcome field allows, then further details about the fate of hostages/kidnapped may be recorded in the "Additional Notes" field.

## XVIII. SOURCE INFORMATION

*A. First Source Citation (scite1) Text Variable*

This field cites the first source used to compile information on the specific incident.

*B. Second Source Citation (scite2) Text Variable*

This field cites the first source used to compile information on the specific incident.

*C. Third Source Citation (scite3) Text Variable*

This field cites the first source used to compile information on the specific incident.

## XIX. DATA MINING TECHNIQUES CONSIDERED

*A. Traditional data mining techniques such as association analysis, classification and prediction, cluster analysis, and outlier analysis identify patterns in structured data [5] and hence these techniques are applicable in this scenario as well. Various forms of data mining especially, crime / terror data mining raises privacy concerns [6]. Nevertheless, researchers are developing various automated data mining techniques for both local law enforcement and national security applications.*

*B. Clustering techniques can group the above data items into classes with similar characteristics to maximize or minimize intraclass similarity—for example, to identify perpetrators who claimed to have carried out the incident in similar ways or distinguish among groups belonging to different terrorist outfits. These techniques do not have a set of pre-defined classes for assigning items. However since the data under question is a Global Data a great deal of presummarisation is involved and hence it would be more apt to have multi dimensional cubes generated and explore/ evolve data mining techniques suited for OLAP. So we do not propose direct application of this method.*

*C. To predict terrorist activity trends, classification can reduce the time required to identify the perpetrators. However, the technique requires a predefined classification scheme[13]. We can evolve a scheme but then classification also requires reasonably complete training and testing data because a high degree of missing data would limit prediction accuracy. The GTD data is sparse in this regard and hence this approach prima facie does not appear promising and hence we do not propose to use this*

*D. With association rule mining we can discover frequently occurring item sets in the GTD database and present the patterns as rules. We can apply this technique to the incidents and perpetrators to help detect potential future incidents of similar nature [12].*

*E. Similar to association rule mining, we shall also try sequential pattern mining to find frequently occurring sequences of incidents that occurred at different times. This approach can identify attack patterns among time-stamped data. Showing hidden patterns benefits terror incidence analysis, but to obtain meaningful results GTD which is a feature rich data has to be summarised and highly structured for which we propose to constuct relevant OLAP cubes for analysis and data mining.*

*F. On these OLAP cubes we shall also be trying out deviation detection / outlier analysis by appyling our own uses specific measures in the form of outlier score functions to study incident data that differs markedly from the rest of the data.*

## XX. ACKNOWLEDGMENT

*A. The authors acknowledge the efforts being made by researchers and Journalists all over the world who are compiling the Global Terror Data Base. The Text, Introduction to Data Mining with Case Studies By G.K. Gupta [1] presents a rich collection of Data Mining Case Studies which has motivated us to explore application of suitable data mining methods to GTD.*

## XXI. REFERENCES

[1] *Introduction to Data Mining with Case Studies By G.K. Gupta*

[2] *http://www.start.umd.edu/gtd/contact/*

[3] *W. Chang et al., "An International Perspective on Fighting Cybercrime," Proc. 1st NSF/NIJ Symp. Intelligence and Security Informatics, LNCS 2665, Springer-Verlag, 2003.*

[4] *http://www.start.umd.edu/gtd/downloads/Codebook.pdf*

[5] *J. Han and M. Kamber, Data Mining: Concepts and Techniques, Morgan Kaufmann, 2001.*

[6] *H. Kargupta, K. Liu, and J. Ryan, "Privacy-Sensitive Distributed Data Mining from Multi-Party Data," Proc. 1st NSF/NIJ Symp. Intelligence and Security Informatics, LNCS 2665, Springer-Verlag, 2003.*

[7] *M.Chau, J.J. Xu, and H. Chen, "Extracting Meaningful Entities from Police Narrative Reports, Proc. Nat'l Conf. Digital Government Research, Digital Government Research Center, 2002.*

[8] *A. Gray, P. Sallis, and S. MacDonell, "Software Forensics: Extending Authorship Analysis Techniques to Computer Programs," Proc. 3rd Biannual Conf.*

[9] *Int'l Assoc. Forensic Linguistics, Int'l Assoc. Foren sic Linguistics, 1997.*

[10] *R.V. Hauck et al., "Using Coplink to Analyze Criminal-Justice Data," Computer, Mar. 2002.*

[11] *T. Senator et al., "The FinCEN Artificial Intelligence System: Identifying Potential Money Laundering from Reports of Large Cash Transactions," AI Magazine, vol.16, no. 4, 1995.*

[12] *.W. Lee, S.J. Stolfo, and W. Mok, "A Data Mining Framework for Building Intrusion Detection Models," Proc. 1999 IEEE Symp. Security and Privacy, IEEE CS Press, 1999.*

[13] *O. de Vel et al., "Mining E-Mail Content for Author Identification Forensics," SIGMOD Record, vol. 30, no. 4, 2001.*

[14] *.G. Wang, H. Chen, and H. Atabakhsh, "Automatically Detecting Deceptive Criminal Identities," Comm. ACM, Mar. 2004.*